\documentclass[a4paper,11pt]{article}
\usepackage{pos}
\usepackage{subcaption}
\usepackage{float}
\usepackage{titlesec}
\titlespacing{\section}{0pt}{*3}{*1}

\title{Event-by-event investigation of the kaon pair-source function with EPOS}

\author*[a]{László Kovács}
\author[a]{Dániel Kincses}
\author[a]{Máté Csanád}

\affiliation[a]{ELTE Eötvös Loránd University,\\
  Pázmány Péter sétány 1/A, Budapest, Hungary}

\emailAdd{laszlokov@staff.elte.hu}
\emailAdd{kincses@ttk.elte.hu}
\emailAdd{csanad@elte.hu}

\abstract{In high-energy collisions, we can obtain information about the source function by measuring the two-particle Bose-Einstein correlation function and considering its relationship with the phase-space density of the particle-emitting source. While a Gaussian shape is commonly assumed, measurements and anomalous diffusion suggest Lévy-stable distributions, as observed in the PHENIX experiment for kaon-kaon pair-source functions. Event generators like EPOS allow direct investigation of freeze-out coordinates, facilitating the analysis of the source function. EPOS, a Monte Carlo-based model, simulates high-energy nuclear and particle collisions, integrating Parton-Based Gribov-Regge theory for initial evolution, subsequent hydrodynamic evolution, and hadronization. In this paper, we present an event-by-event analysis of the kaon source function in $\sqrt{s_\text{NN}}$ = 200 GeV Au+Au collisions using the EPOS model.}

\FullConference{42nd International Conference on High Energy Physics (ICHEP2024)\\
18-24 July 2024\\
Prague, Czech Republic\\}

\begin{document}
\maketitle

\section{Theoretical background}
Femtoscopy is a widely used method to study the space-time geometry of the particle-emitting source created in heavy-ion collisions~\cite{FEM}. This field of high-energy nuclear and particle physics investigates correlations of particle pairs, allowing us to probe the quark-gluon plasma on the femtometer scale. Our primary focus is on studying the shape of the phase-space density of the particle-emitting source.  While a Gaussian shape is reasonable according to the central limit theorem, a more general approach can be adopted. The spherically symmetric, Lévy-stable distribution can be defined as
\begin{equation}
\mathcal{L}(\textbf{r};R,\alpha) = \frac{1}{(2\pi)^3} \int \text{d}^3 \textbf{q} \:  e^{i\textbf{q}\textbf{r}} e^{-\frac{1}{2}|\textbf{q}R|^\alpha},
\end{equation}
where $R$ is the Lévy scale parameter, $\alpha$ is the Lévy exponent, and $\textbf{q}$ is a three-dimensional integration variable. The parameter $\alpha$ characterizes the shape of the distribution: when $\alpha = 2$, it is Gaussian, while in the case of $\alpha < 2$, it follows a power-law behavior.

A Lévy-shaped source can arise from various factors.  If anomalous diffusion is the underlying cause, we expect higher $\alpha$ values for pions compared to kaons~\cite{ANO}. Experimental studies have demonstrated that Lévy-stable source distributions in $\sqrt{s_{\text{NN}}}=200$ GeV Au+Au collisions provide a high-quality, statistically valid description of the measured correlation functions for pions~\cite{PION} and kaons~\cite{KAON}. The pair source distribution (which can be reconstructed indirectly from femtoscopic momentum correlation measurements) is defined as the autocorrelation of the single-particle phase-space density $S(x, p)$:
\begin{equation}
D(r,K) = \int S \left(\rho + \frac{r}{2},K \right) S \left(\rho -\frac{r}{2},K \right) d^4 \rho.
\label{eq:D}
\end{equation}
Here, the following variables are used: the pair center-of-mass four-vector $\rho = (x_1 + x_2)/2$, the pair separation four-vector $r = x_1 - x_2$, and the average momentum $K = (p_1 + p_2)/2$.

\section{Analysis}
To gain a deeper understanding of the processes underlying the experimental results, further effort is required from the phenomenology side. Key tools for these investigations include event generators, which integrate various theoretical and phenomenological methods to model nuclear reactions. One such event generator is the EPOS model~\cite{EPOS} -- the \textbf{E}nergy conserving quantum mechanical multiple scattering approach, based on \textbf{P}artons (parton ladders), \textbf{O}ff-shell remnants, and \textbf{S}plitting of parton ladders. It utilizes Monte Carlo techniques and was developed to describe processes occurring in collisions at $\mu_B \approx 0$, at very high energies, and for various systems, including Au+Au, Pb+Pb, and p+p. 

For the analysis, events were generated using EPOS359. By utilizing like-sign kaon pairs, the one-dimensional pair-source distribution was measured in the longitudinal co-moving system (LCMS). The $D(r_{\text{LCMS}})$ histogram, representing the pair-source distribution after angle- and time-integration, was constructed by counting the number of pairs within each $r_{\text{LCMS}}$ bin, where $r_{\text{LCMS}}$ is the one-dimensional pair-separation vector. Following an event-by-event analysis, the mean and standard deviation of the Lévy parameters were obtained from thousands of fits, similarly to Ref.~\cite{EPOSPION}. The measurement was carried out for five different average transverse mass ($m_\text{T}$) ranges and four centrality classes. 

\section{Results}
\begin{figure}[H]
    \centering
    \begin{subfigure}{0.44\textwidth}
        \includegraphics[width=\textwidth]{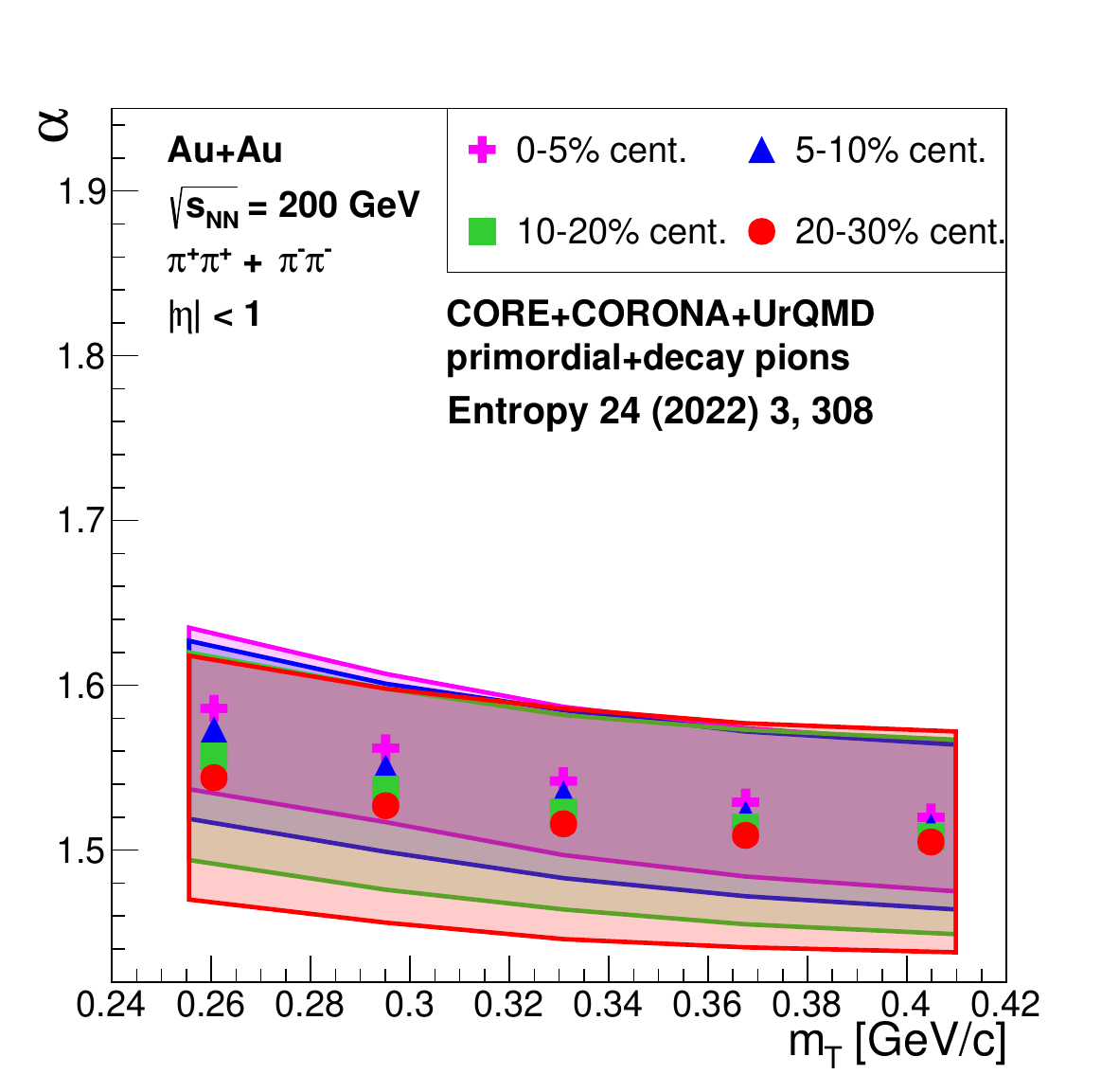}   
    \end{subfigure}
    \hspace{0.02\textwidth}  
    \begin{subfigure}{0.44\textwidth}
        \includegraphics[width=\textwidth]{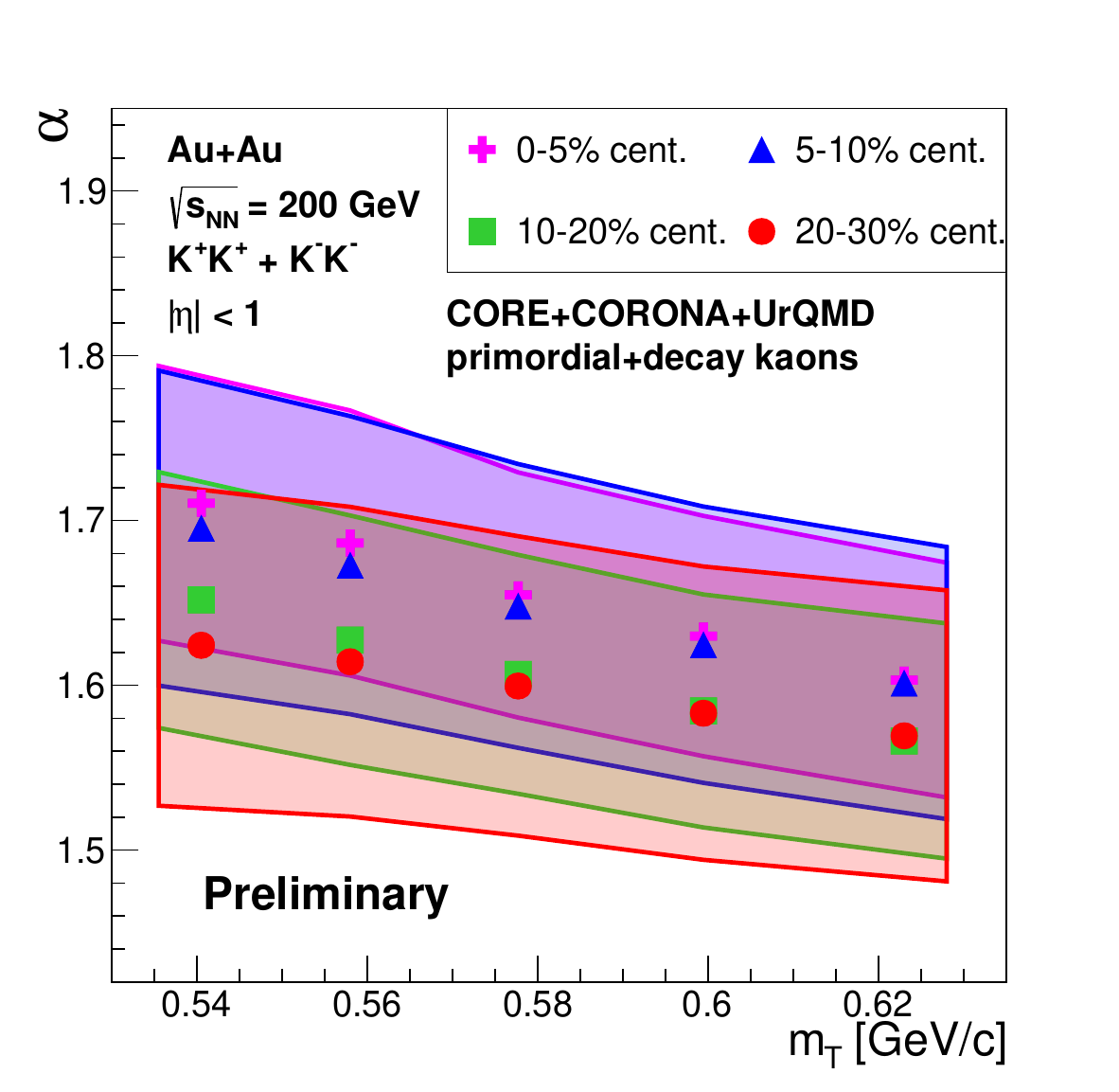}   
    \end{subfigure}
    \caption{The mean $\alpha$ values are shown as a function of transverse mass for four different centrality classes. The colored boxes represent the standard deviation. The plot on the left displays the results for pions, while the plot on the right presents the kaon results.}
    \label{fig:RES}
\end{figure}

The results for the kaon $\alpha$ values are displayed on the right side of Fig.~\ref{fig:RES}, while the pion $\alpha$ values, for reference, are shown on the left side, as reported in Ref.~\cite{EPOSPION}. These results indicate that the Lévy exponent $\alpha$ for kaons is larger than that for pions. This trend is the opposite of what is expected in the case of anomalous diffusion~\cite{ANO}. This discrepancy suggests that additional factors beyond anomalous diffusion may influence the source distribution, calling for further investigation into the underlying dynamics.

\acknowledgments{This research was funded by the NKFIH grants TKP2021-NKTA-64, PD-146589, K-146913, and K-138136. D. K. and L. K. were also supported by the EKÖP-24 University Excellence Scholarship program of the Ministry for Culture and Innovation from the source of the national research, development, and innovation fund.}

\end{document}